\documentclass[aps,prl,twocolumn,superscriptaddress,citeautoscript,longbibliography]{revtex4-2}
\usepackage[utf8]{inputenc}
\usepackage{multirow}
\usepackage{hhline}
\usepackage{enumerate}
\usepackage{amssymb,bm}
\usepackage{graphicx}
\usepackage{amsmath}
\usepackage{braket}
\usepackage{mathtools}
\usepackage{bbm}
\usepackage{booktabs}
\usepackage{tabularx}
\usepackage[colorlinks=true,linkcolor=blue,citecolor=blue,urlcolor=blue]{hyperref}
\usepackage{xcolor,soul}
\usepackage{changes}
\usepackage{amsthm}
\usepackage{mathrsfs}
\usepackage{subfigure}
\usepackage{tikz}
\usetikzlibrary{calc}

\newtheorem{theorem}{Theorem}[section]

\newcommand{\Tr}{\operatorname{Tr}}

\newcommand{\iu}{{i\mkern1mu}}
\newcommand{\diff}[1]{\mathop{}\!\mathrm{d}#1}

\newcommand{\Hbare}{H_0}
\newcommand{\HZ}{H_Z}

\newcommand{\linOpCdk}[1]{\mathcal{L}((\bC^{d})^{\otimes #1 k})}
\newcommand{\Mk}[1]{M_{#1}^{(k)}}

\newcommand{\Pslow}{\mathcal{P}_{\text{s}}}

\newcommand{\bE}{\mathbb{E}}
\newcommand{\bC}{\mathbb{C}}

\newcommand{\calL}{\mathcal{L}}

\newcommand{\calK}{\mathcal{K}}

\newcommand{\calT}{\mathcal{T}}

\newcommand{\calU}{\mathcal{U}}
\newcommand{\id}{\mathbbm{1}}
\newcommand{\idk}[1]{\mathbbm{1}^{\otimes #1 k}}

\newcommand{\GHZ}{\text{GHZ}}

\newcommand{\kket}[1]{|{#1}\rangle\!\rangle}
\newcommand{\bbra}[1]{\langle\!\langle {#1}|}

\begin{document}

	\preprint{APS/123-QED}

	\title{Optimal Disorder Strength for Approximate Unitary $k$-Designs in Spin Chains}

    \author{Tian-Gang Zhou}
    \email{tgzhou.physics@gmail.com}
    \thanks{These authors contributed equally to this work.}
\affiliation{DQMP, University of Geneva, 24 quai Ernest-Ansermet, 1211 Genève 4, Switzerland}
\affiliation{Geneva Quantum Center, University of Geneva, 24 quai Ernest-Ansermet, 1211 Genève 4, Switzerland}

	\author{Yi-Neng Zhou}
\thanks{These authors contributed equally to this work.}
\affiliation{Department of Theoretical Physics, University of Geneva, 24 quai Ernest-Ansermet, 1211 Genève 4, Switzerland}
\affiliation{Geneva Quantum Center, University of Geneva, 24 quai Ernest-Ansermet, 1211 Genève 4, Switzerland}

	\author{Thierry Giamarchi}
    \email{Thierry.Giamarchi@unige.ch}
	\affiliation{DQMP, University of Geneva, 24 quai Ernest-Ansermet, 1211 Genève 4, Switzerland}
    \affiliation{Geneva Quantum Center, University of Geneva, 24 quai Ernest-Ansermet, 1211 Genève 4, Switzerland}

	\author{Julian Sonner}
\email{julian.sonner@unige.ch}
	\affiliation{Department of Theoretical Physics, University of Geneva, 24 quai Ernest-Ansermet, 1211 Genève 4, Switzerland}
\affiliation{Geneva Quantum Center, University of Geneva, 24 quai Ernest-Ansermet, 1211 Genève 4, Switzerland}

	\date{\today}

	\begin{abstract}

Quantum simulators with disordered spin Hamiltonians offer a natural route to Haar-like unitary ensembles, but face a competition. Disorder enhances randomness, yet strong disorder suppresses dynamics and localizes the system in the computational basis. This suggests an optimal intermediate disorder strength for approximating a unitary $k$-design, which has so far been identified mainly empirically. Here, we develop a perturbative theory for the moment gap $\Delta^{(k)}$, the deviation of the $k$-th moment operator from its Haar value, for the full spectrum of the interacting XXZ chain in a fixed-magnetization sector. We find the non-monotonic asymptotic scalings $\ln \Delta^{(k)}\sim -\sigma^2$ at weak disorder and $\ln \Delta^{(k)}\sim -1/\sigma$ at strong disorder, whose crossover sets an optimal disorder strength $\sigma_c\sim J_{xy}$, where $J_{xy}$ is the energy scale of the non-random interaction. We further show that a multi-quench protocol further suppresses the gap exponentially in the number of quenches. Numerical simulations confirm both scalings and this exponential suppression, and we demonstrate the usefulness of the protocol for classical shadow tomography. This work suggests that strong disorder may block both thermalization and the formation of unitary designs through a common localization mechanism.

	\end{abstract}

	\maketitle

\paragraph{\color{blue}Introduction---}\label{sec:intro}
 Unitary $k$-designs~\cite{ambainis2007quantumtdesignstwiseindependence,dankert2005efficientsimulationrandomquantum,Dankert2009PRA,Mele2024introductiontohaar} are ensembles of random unitaries whose first $k$ moments match those of the Haar ensemble. They are essential resources in quantum information science, underpinning protocols such as classical shadow tomography~\cite{Huang_2020,Elben_2022}, randomized measurements and benchmarking~\cite{Emerson_2005,Knill_2008,Elben_2019,Vermersch_2019}, and studies of deep thermalization~\cite{IppolitiHo2022Quantum,lami2025quantumstatedesignemergent,Cotler_2017} and quantum information scrambling~\cite{Sekino_2008,Hayden_2006,Roberts_2017}. Theoretical results guarantee that local random circuits generate approximate $k$-designs in polynomial depth~\cite{Brandao2016,Haferkamp2022Quantum,Harrow_2009,HarrowLow2009ApproxRandom,Brand_o_2021,laracuente2024approximateunitarykdesignsshallow,schuster2025strongrandomunitariesfast}, motivating implementations on near-term quantum hardware. Yet achieving sufficiently random unitaries on analog quantum simulators, which are confined to a fixed Hamiltonian structure~\cite{cui2025randomunitarieshamiltoniandynamics,zhou2026hamiltonianssufficientunitarykdesign,zhou2025realizingunitarykdesignssingle,zhou2026solvablerandomunitarydynamics}, remains a fundamental challenge.

Physical quantum simulators realize one-dimensional spin chains and naturally support random on-site potentials drawn from engineered disorder \cite{Anderson1958Absence,Billy2008Anderson, EversMirlin2008Anderson,Fallani2007BoseGlass,Schreiber2015MBL,GiamarchiSchulz1988Anderson}. Representative platforms include Rydberg atom arrays~\cite{Signoles:2019us,Joshi_2020}, cold-atom optical lattices~\cite{Bloch:2012ee,smaleObservationTransitionDynamical2019}, dipolar molecules~\cite{Yan:2013fn,Hazzard:2014bx}, and NV-center systems~\cite{Lukin17NV,Lukin18NV}. In combination with randomized-measurement protocols~\cite{Huang_2020,Vermersch_2018,Elben_2022}, time evolution under such disordered Hamiltonians can itself provide the randomness required to generate approximate unitary designs~\cite{zhou2025realizingunitarykdesignssingle,zhou2026hamiltonianssufficientunitarykdesign,sun2026unitarydesignschaotichamiltonians}. The resulting ensemble is governed by a competition: increasing disorder enhances dephasing and randomness, while strong disorder suppresses dynamical mixing and pins the system near the computational basis. This suggests an optimal intermediate disorder strength for approximating a unitary $k$-design. However, an analytic determination of this optimum from $k$-design convergence has been lacking.

    In this work, we determine the optimal disorder strength for generating unitary $k$-designs from disordered spin-chain dynamics and show how a multi-quench protocol exploits this optimum to efficiently produce approximate $k$-designs. Our analysis develops a replica-operator expansion for the spectral gap $\Delta^{(k)}$ between the $k$-th moment operator generated by disordered spin-chain evolution and its Haar counterpart, valid for arbitrary $k$ and the full spectrum of the interacting XXZ chain within a fixed-magnetization sector. We find the non-monotonic scalings $\ln \Delta^{(k)}\sim -\sigma^2$ at weak disorder and $\ln \Delta^{(k)}\sim -1/\sigma$ at strong disorder, whose crossover determines an optimal disorder strength $\sigma_c\sim J_{xy}$. Moreover, the multi-quench protocol suppresses the gap as $\Delta_q^{(k)}\leq(\Delta^{(k)})^{q+1}$, leading to exponentially fast convergence toward the Haar ensemble with the number of quenches. Direct numerical simulations of the XXZ chain confirm both perturbative expressions and the scalings, and we demonstrate a practical realization of a unitary $2$-design for classical-shadow tomography~\cite{Huang_2020,aaronson2018shadow, struchalin2021experimental,Elben_2022, hu2023classical}.

\paragraph{\color{blue} Model and metric---}\label{sec:model} We consider a one-dimensional spin chain with disorder $H=H_0+H_Z$, where $H_0$ is the XXZ Hamiltonian~\cite{Yang1967,Luther1975,Giamarchi2004},
\begin{equation}\label{eq:HXXZ}
    H_0 = \sum_{i=1}^{L-1} \left[ J_{xy}\left(X_iX_{i+1} + Y_i Y_{i+1}\right) + J_z Z_iZ_{i+1} \right] ,
\end{equation}
with $X_i, Y_i, Z_i$ the Pauli operators on site $i$ and open boundary conditions in the numerics. The random longitudinal field is $H_Z=\sum_{j=1}^{L} h_j Z_j$, with independent Gaussian on-site fields $h_j^{(a)}\sim\mathcal N(0,\sigma^2)$, i.e., $\mathbb E[h_i^{(a)}h_j^{(b)}]=\sigma^2\delta_{ij}\delta_{ab}$, where $a$ labels the quench and $\sigma$ sets the disorder strength. We set $J_{xy}=1$ and fix $J_z$ to representative values of the same order, so that $\sigma$ is the only tuning parameter. Since $H_0$ and $\HZ$ both commute with $S^z=\sum_i Z_i$, we work in the $S^z=0$ sector, and omit global rotations between quenches to preserve it.

	The closeness of the random ensemble $\nu$ generated by Hamiltonian time evolution to the Haar ensemble $\mu_H$ can be quantified through their $k$-th moment operators on the doubled $k$-replica Hilbert space~\cite{Mele2024introductiontohaar,collinsIntegrationRespectHaar2006a},
\begin{equation}
\Mk{\nu} =
\underset{U\sim\nu}{\bE}
\left[ U^{\otimes k}\otimes U^{*\otimes k} \right].
\end{equation}
Here, $U\in\calL(\bC^d)$ acts within a conserved subspace of dimension $d$, such that $\Mk{\nu}\in\linOpCdk{2}$, where $\calL$ means the linear operator. For the Haar ensemble, $\Mk{\mu_H}$ projects onto the commutant of $U(d)^{\otimes k}$ and admits the Weingarten-coefficient representation reviewed in Sec.~\ref{app:haar_basis} of the Supplemental Material (SM)~\cite{SM}. We quantify the deviation of $\nu$ from a unitary $k$-design by the spectral-norm distance between its moment operator and the Haar moment operator,
\begin{equation}
\label{eq:gap_def}
\Delta^{(k)} = \big\| \Mk{\nu} - \Mk{\mu_H} \big\|_{\infty},
\end{equation}
where $\|\cdot\|_\infty$ denotes the largest singular value.

A complementary diagnostic is the frame potential (FP)~\cite{Roberts_2017,Cotler_2017,Brown_2018,Gross_2007,Roy_2009}, whose excess over the Haar value equals $\|\Mk{\nu}-\Mk{\mu_H}\|_2^2\leq d^{2k}(\Delta^{(k)})^2$, where $\|\cdot\|_2$ is the Hilbert--Schmidt norm. Since it receives contributions from the full singular-value spectrum, the FP can show richer time dependence than the moment gap, which depends only on the largest singular value. For $k=1$, we show in Sec.~\ref{app:fp_asymptotics} of the SM~\cite{SM} that the FP gap, normalized by its disorder-limit plateau, has the same leading $-\sigma^2$ and $-\sigma^{-1}$ log-scalings as the moment gap in the weak- and strong-disorder regimes, respectively. However, these plateaus differ, $d^2-1$ and $d-1$, which complicates a uniform perturbative treatment. We therefore use the moment gap as our main diagnostic.

These diagnostics characterize the unitary ensemble over the entire $S^z=0$ sector and contain no finite temperature restriction to a narrow energy window. In particular, normalized traces $d^{-1}\Tr(\cdot)$ correspond to expectation values in the maximally mixed state, i.e., at infinite temperature within this sector. This is different from the low-temperature probe of the unitary $k$-design~\cite{zhou2026solvablerandomunitarydynamics}.

\paragraph{\color{blue} Perturbative analysis of the moment gap---}\label{sec:perturbation} We determine the optimal disorder strength for approximate unitary $k$-designs by analyzing the moment gap $\Delta^{(k)}$ in the weak- and strong-disorder regimes. Intuitively, weak disorder scrambles the Hilbert space insufficiently, while strong disorder pins the spins along $\hat z$ and yields a basis-locked ensemble~\cite{Anderson1958Absence,MBL1,MBL2} far from the Haar ensemble, so an optimum must lie near $\sigma\sim J_{xy}$. We first consider the single-quench case, where the evolution is governed by a single Hamiltonian.

For weak disorder, $\sigma \ll J_{xy}$, we treat $\HZ$ perturbatively. In the interaction picture of $\Hbare$,
\begin{equation}\label{eq:smallsigma_IP}
    U(t) = e^{-\iu \Hbare t}\, \mathcal{T}\exp\!\left(-\iu\int_0^t\! \diff{s}\, \HZ(s)\right),
\end{equation}
with $\HZ(s) = \sum_i h_i Z_i(s)$ and $Z_i(s) = e^{\iu \Hbare s}Z_i e^{-\iu \Hbare s}$. 

In the doubled replica space, we write $\calU^{k,k}\equiv U^{\otimes k} \otimes U^{*\otimes k}$ and, for any single-copy operator $O$, define its replica counterpart $O^{k,k} = \sum_{\alpha=1}^{k} O^{(\alpha)} - \sum_{\beta=1}^{k} O^{*(\beta+k)}$, where $\alpha$ ($\beta$) labels forward (backward) replicas, the minus sign and complex conjugation reflect the backward evolution under $U^*$. Equation~\eqref{eq:smallsigma_IP} then becomes $\calU^{k,k} = \calU_0^{k,k}\,\calT\exp[-\iu \sum_j h_j \int_0^t \diff{s}\, Z_j^{k,k}(s)]$, with $\calU_0^{k,k} = e^{-\iu H_0^{k,k} t}$ and $Z_j^{k,k}(s)=e^{\iu H_0^{k,k}s}Z_j^{k,k}e^{-\iu H_0^{k,k}s}$. Averaging over the Gaussian fields gives $\Mk{\nu} = \calU_0^{k,k}\, \Mk{I,\nu}$ with
\begin{equation}\label{eq:replica_M_smallsigma}
    \Mk{I,\nu} = \calT \exp\!\left( -\tfrac{1}{2}\sigma^2 \sum_j \int_{0}^{t}\!\!\int_{0}^{t}\! \diff{s}\diff{s'}\, Z_j^{k,k}(s) Z_j^{k,k}(s')\right).
\end{equation}

Using the norm identity $\|\calU_0^{k,k} A\|_\infty=\|A\|_\infty$ and the Haar random property, the gap reduces to
\begin{equation}\label{eq:spectral_gap_proj}
    \Delta^{(k)} = \big\| \Mk{\mu_H,\perp} \Mk{I,\nu} \Mk{\mu_H,\perp} \big\|_{\infty},
\end{equation}
with $\Mk{\mu_H,\perp} = \idk{2} - \Mk{\mu_H}$. Expanding to second order in $\sigma$, which is controlled when $\sigma^2$ times the relevant eigenvalue of $\Omega^{(k)}$ is small, $\Mk{I,\nu} \approx \idk{2} - \sigma^2 \Omega^{(k)}(t)$ with
\begin{equation}
    \Omega^{(k)}(t) = \sum_i \int_{0}^{t}\!\!\int_{0}^{s}\! \diff{s}\diff{s'}\, Z_i^{k,k}(s) Z_i^{k,k}(s'),
\end{equation}
and identifying the smallest non-zero singular value gives
\begin{equation}\label{eq:gap_smallsigma}
    \begin{aligned}
    \Delta^{(k)}(t) ={}& 1 - \tfrac{\sigma^2}{2} \lambda_{\min}\!\left[ \Mk{\mu_H,\perp}\!\left(\Omega^{(k)} + \mathrm{h.c.}\right)\!\Mk{\mu_H,\perp} \right] \\
    &{}+ \mathcal{O}(\sigma^4),
    \end{aligned}
\end{equation}
where $\lambda_{\min}$ denotes the smallest eigenvalue restricted to the range of $\Mk{\mu_H,\perp}$.
For the second term, the single-replica case reduces this minimization to the smallest time-integrated squared commutator $\sum_i d^{-1}\|[\int_0^t Z_i(s)\diff{s},O]\|_2^2$, an out-of-time-commutator, over traceless normalized operators $O$, which connects to quantum scrambling~\cite{maldacenaBoundChaos2016,MaldacenaStanford2016PRD,robertsLocalizedShocks2015a,shenkerBlackHolesButterfly2014,shenkerStringyEffectsScrambling2015b}. The larger squared commutator means the smaller moment gap, which agrees with the definition of the randomness. Since $\Omega^{(k)}$ is independent of $\sigma$, the leading correction to the logarithm of the gap scales as
\begin{equation}\label{eq:scaling_smallsigma}
    \ln \Delta^{(k)} \sim -\sigma^2.
\end{equation}

\begin{figure}
    \centering
    \includegraphics[width=1.0\linewidth]{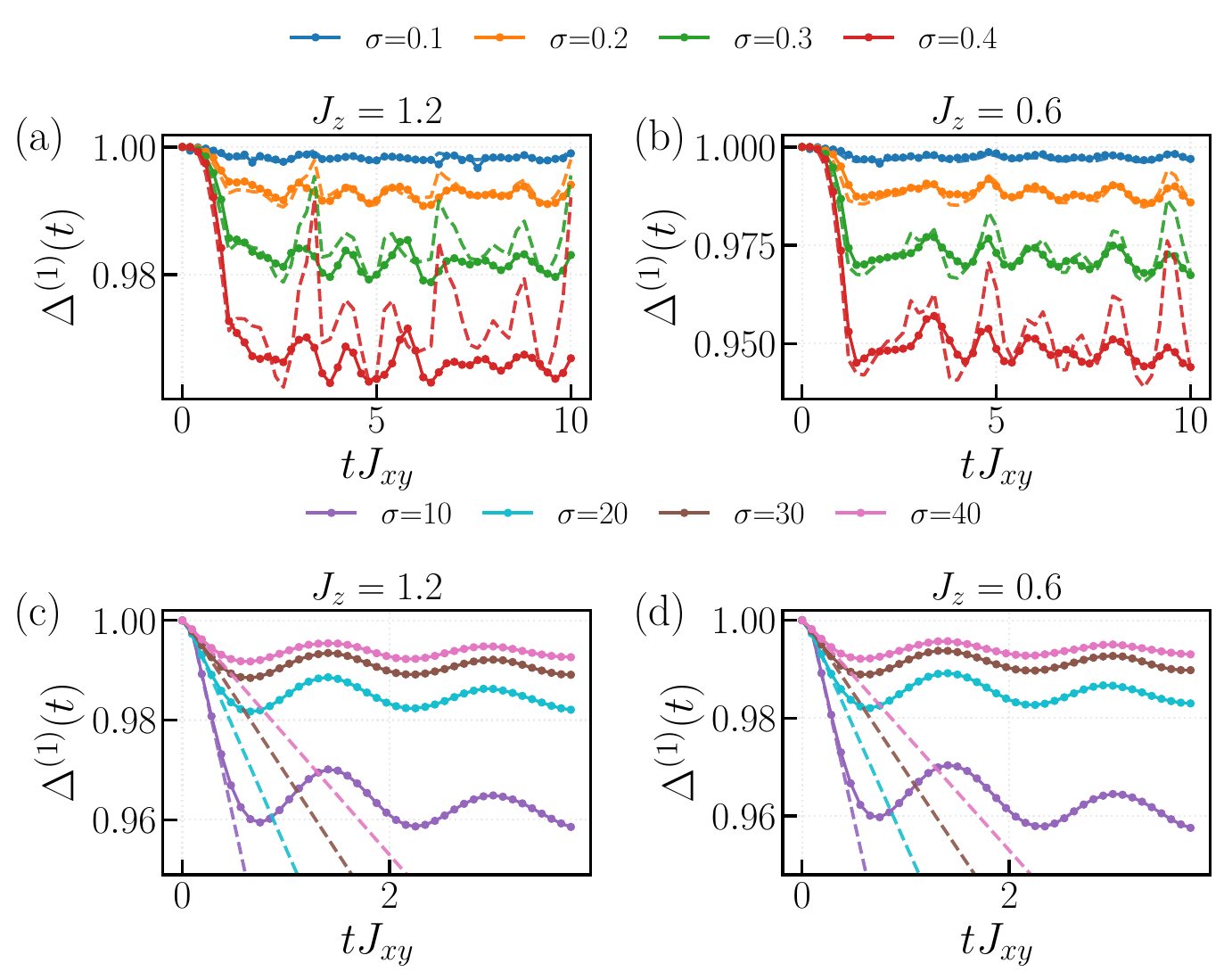}
    \caption{Disorder-averaged first-moment gap $\Delta^{(1)}(t)$ of an open XXZ chain with $L=6$ and $J_{xy}=1$, restricted to the $S^z=0$ sector. Solid lines with markers show Monte Carlo results and dashed lines show the perturbative predictions. (a,b) Weak disorder, $\sigma=0.1,0.2,0.3,0.4$, for $J_z=1.2$ and $0.6$, respectively, compared with Eq.~\eqref{eq:gap_smallsigma}. (c,d) Strong disorder, $\sigma=10,20,30,40$, for the same two values of $J_z$, compared with Eq.~\eqref{eq:gap_largesigma}. The leading strong-disorder decay is independent of $J_z$. At later times, the numerical gaps fluctuate around nonzero plateaus, whereas the strong-disorder Markovian prediction continues to decrease.}
    \label{fig:verify_xxz}
\end{figure}

In the opposite, strong-disorder limit $\sigma\gg J_{xy}$, the random fields suppress transitions between $Z$-basis configurations, and we treat $\Hbare$
perturbatively in the interaction picture of $\HZ$:
\begin{equation}\label{eq:largesigma_IP}
    U(t)
    =
    e^{-\iu \HZ t}
    \mathcal{T}
    \exp\!\left[
        -\iu\int_0^t \diff{s}\,\Hbare(s)
    \right],
\end{equation}
where $\Hbare(s)=e^{\iu\HZ s}\Hbare e^{-\iu\HZ s}$. Both $e^{-iH_Z t}$ and $H_0(s)=e^{iH_Zs}H_0e^{-iH_Zs}$ depend on the same disorder realization $\{h_j\}$. Their disorder averages are thus correlated and cannot be evaluated independently in general.

For $\sigma t\gg1$, disorder-induced dephasing instead selects the common zero-eigenspace of the local replica-imbalance operators $Z_i^{k,k}$. To characterize this slow subspace, let
$Z_i\ket{j}=z_i(j)\ket{j}$, with $z_i(j)=\pm1$, for the
computational basis in the chosen $d$-dimensional sector.
For a replica configuration
$\bm{j}=(j_1,\ldots,j_k)\in[d]^k$, where $[d]=\{1,\ldots,d\}$,
define the local replica sum
$q_i(\bm{j})=\sum_{\alpha=1}^k z_i(j_\alpha)$.

The slow-space projector is
\begin{equation}\label{eq:Pslow_def}
    \Pslow = \sum_{\forall i,\ q_i(\bm{j})=q_i(\bm{\ell})} \ket{\bm{j},\bm{\ell}}\bra{\bm{j},\bm{\ell}},
\end{equation}
where $\ket{\bm{j},\bm{\ell}}=\ket{\bm{j}}_{\mathrm F}\otimes\ket{\bm{\ell}}_{\mathrm B}$. By construction, $Z_i^{k,k}\Pslow=0$ for every site, so $\HZ^{k,k}$ vanishes on this subspace for every disorder realization. The constraint fixes only the replica-summed $Z$ eigenvalue on each site, not its assignment to individual replicas. Hence $\Pslow$ contains all $d^k$ diagonal states $\ket{\bm{j},\bm{j}}$, together with off-diagonal pairs $\bm{j}\neq\bm{\ell}$ such as those related by replica permutations, so that $\operatorname{rank}(\Pslow)\geq d^k$ \footnote{The constraint fixes only the replica sum $q_i$ on each site, so $\bm{\ell}$ may redistribute spins among replicas independently on each site rather than by a global replica permutation. For example, for $k=2$ and $L=4$, $\bm{j}=(\uparrow\downarrow\uparrow\downarrow,\downarrow\uparrow\downarrow\uparrow)$ and $\bm{\ell}=(\uparrow\uparrow\downarrow\downarrow,\downarrow\downarrow\uparrow\uparrow)$ satisfy $q_i(\bm{j})=q_i(\bm{\ell})=0$ on every site, yet $\bm{\ell}$ is not a permutation of $\bm{j}$.}. For fixed $k$ and large $d$, this greatly exceeds $\operatorname{rank}(\Mk{\mu_H})=k!$ (for $d\geq k$). Since $\HZ$ acts trivially on $\Pslow$, it leaves these extra non-Haar modes untouched and alone cannot generate a unitary $k$-design. Reducing the remaining gap thus requires the effective dynamics
induced by $\Hbare$ to relax the additional non-Haar modes within this enlarged slow subspace.

With the similar idea of Mori-Zwanzig formalism\cite{Mori}, projecting the second-order expansion in the $2k$-replica space onto the slow subspace, on which $e^{-\iu H_Z^{k,k}t}$ acts as the identity, gives
\begin{equation}\label{eq:VI_large}
    \Pslow \Mk{\nu} \Pslow =\Pslow \!\left(\idk{2} - \calK^{(k)}(t) \right) \Pslow,
\end{equation}
with the second-order memory kernel
\begin{equation}\label{eq:memory_function}
    \calK^{(k)}(t) =  \int_0^t\!\! \int_{0}^{s}\! \diff{s}\diff{s'}\, \Pslow H_0^{k,k}\, e^{-\iu H_Z^{k,k}(s-s')}\, H_0^{k,k} \Pslow.
\end{equation}
After disorder averaging, the rapidly oscillating phase leads to $\bE\,e^{\iu \sum_i h_i \tilde z_i (s-s')} = e^{-\sigma^2 \sum_i \tilde z_i^2 (s-s')^2/2}$, where $\tilde z_i$ is the eigenvalue of $Z_i^{k,k}$ on the intermediate state, equal to $\pm2$ on the two sites of the hopped bond so that $\sum_i\tilde z_i^2=8$. In the long-time Markovian limit,
\begin{equation}
    \Pslow \calK^{(k)}(t) \Pslow \approx C(\sigma,t)\, \Pslow H_0^{k,k} H_0^{k,k} \Pslow,
\end{equation}
where $C(\sigma,t) = \int_0^t\!\diff{u}\,(t-u)\,e^{-4\sigma^2u^2}\simeq\tfrac{\sqrt{\pi}}{4\sigma}\, t$ for $\sigma t\gg1$. Hence, the moment gap reads,
\begin{equation}\label{eq:gap_largesigma}
    \Delta^{(k)}(t) = 1 - C(\sigma,t) \lambda_{\min}\!\left[ \Mk{\mu_H,\perp}\Pslow\! (H_0^{k,k})^2 \!\Pslow\Mk{\mu_H,\perp} \right].
\end{equation}
 Since $C \propto \sigma^{-1}$, the moment gap in the strong-disorder regime obeys
\begin{equation}\label{eq:scaling_largesigma}
    \ln \Delta^{(k)} \sim -\sigma^{-1}
\end{equation}
for $\sigma\gg J_{xy}$. This result is consistent with many-body localization (MBL)~\cite{MBL1,MBL2,MBL3,MBL4}. Like the localized phase at strong disorder prevents thermalization and ergodicity, strong disorder here also prevents the dynamics from forming a unitary $k$-design.

As Eq.~\eqref{eq:gap_largesigma} shows, for $k=1$ the gap $\Delta^{(1)}$ depends on $H_0$ only through the $X_iX_{i+1}+Y_iY_{i+1}$ couplings. This is because the slow subspace consists of diagonal operators, and the $J_z$ term in Eq.~\eqref{eq:HXXZ} is itself diagonal, so it commutes with every operator in this subspace and therefore annihilates it. Therefore, the first-moment gap should not depend on $J_z$. For $k\geq2$, however, $J_z=0$ is not a valid diagnostic, since the free-fermion chain gives $\Delta^{(k)}=1$ exactly. For $J_z\neq0$, we find numerically that $\Delta^{(2)}$ coincides with $\Delta^{(1)}$ after a short transient, as shown in Sec.~\ref{app:k1k2} of the SM \cite{SM}.

We numerically simulate the XXZ chain and compute the time evolution of the moment gap, as shown in Fig.~\ref{fig:verify_xxz}. In the weak-disorder regime, Fig.~\ref{fig:verify_xxz}(a) and \ref{fig:verify_xxz}(b) show that the results agree well with the perturbative prediction of Eq.~\eqref{eq:gap_smallsigma} at short times and even at longer times with a saturated value. In the strong-disorder regime, Figs.~\ref{fig:verify_xxz}(c) and \ref{fig:verify_xxz}(d) likewise confirm the perturbative prediction of Eq.~\eqref{eq:gap_largesigma} at early times. This approximation cannot work at long times, since the second-order Markovian prediction keeps decreasing linearly in $t$, whereas the static disorder eventually leads to saturation, where the formula works until that time. Moreover, the $\Delta^{(1)}$ is independent of $J_z$, in agreement with the analysis above.

The non-monotonic minimum implied by Eqs.~\eqref{eq:scaling_smallsigma} and \eqref{eq:scaling_largesigma} sits at $\sigma_c \sim J_{xy}$, where both perturbative expansions break down. However, the single-quench protocol cannot reach Haar randomness, because at late times a single Hamiltonian only dephases the ensemble in its own eigenbasis, so $\Delta^{(k)}$ saturates at a nonzero plateau.

\paragraph{\color{blue}Multi-quench protocol---}\label{sec:multiplequench}
To further suppress the gap  with extra quenches, driving the ensemble toward a unitary $k$-design, we consider a multi-quench protocol that contains evolutions under independently sampled disorder realizations~\cite{Sen2018ZigzagTPE,zhou2025realizingunitarykdesignssingle},
\begin{equation}\label{eq:U_q_quench}
    U(t) = \prod_{a=1}^{q+1} e^{-\iu H^{(a)} \Delta t_a},
\end{equation}
where $a=1,\dots,q+1$ labels the disorder realizations and $q$ is the number of switches between Hamiltonians. The switching times $t_1,\ldots,t_q$, together with $t_0=0$ and $t_{q+1}=t$, partition the total evolution into $q+1$ intervals of duration $\Delta t_a=t_a-t_{a-1}$. The clean part $H_0$ is fixed across quenches, and $q=0$ recovers the single-quench limit.

For independent disorder realizations the moment operator factorizes,
$\Mk{\nu_q}=\Mk{\nu}(\Delta t_{q+1})\cdots\Mk{\nu}(\Delta t_1)$. Since
$\Mk{\nu}\Mk{\mu_H}=\Mk{\mu_H}\Mk{\nu}=\Mk{\mu_H}$, one has exactly
$\Mk{\nu_q}-\Mk{\mu_H}=\prod_{a}\big[\Mk{\nu}(\Delta t_a)-\Mk{\mu_H}\big]$, and
submultiplicativity of the spectral norm gives the non-perturbative bound
\begin{equation}\label{eq:gap_q_bound}
  \Delta_q^{(k)}\le\prod_{a=1}^{q+1}\Delta^{(k)}(\Delta t_a),
\end{equation}
i.e., $\Delta_q^{(k)}\le(\Delta^{(k)})^{q+1}$ for equal segments. The bound is
approximately saturated when all segments share their slowest non-Haar mode. At
strong disorder, $e^{-\iu H_Z^{k,k}t}$ acts trivially on $\Pslow$, and in the
Markovian limit each segment contributes
$\idk{2}-C(\sigma,\Delta t_a)\Pslow(H_0^{k,k})^2\Pslow$. All factors are
functions of one Hermitian operator, giving
$\Delta_q^{(k)}\approx\prod_a[1-C(\sigma,\Delta t_a)\lambda_{\min}]$. At weak
disorder the factors are $\Mk{I,\nu}$ conjugated by the free evolution up to each
switching time and are not Hermitian, but the bound is still nearly saturated if the slowest non-Haar mode $\kket{O}$ is conserved by the clean dynamics, $H_0^{k,k}\kket{O}=0$. The free evolution between switches then leaves $\kket{O}$ unchanged, so every segment suppresses it by approximately the same factor $\Delta^{(k)}$.

\begin{figure}
    \centering
\includegraphics[width=1.0\linewidth]{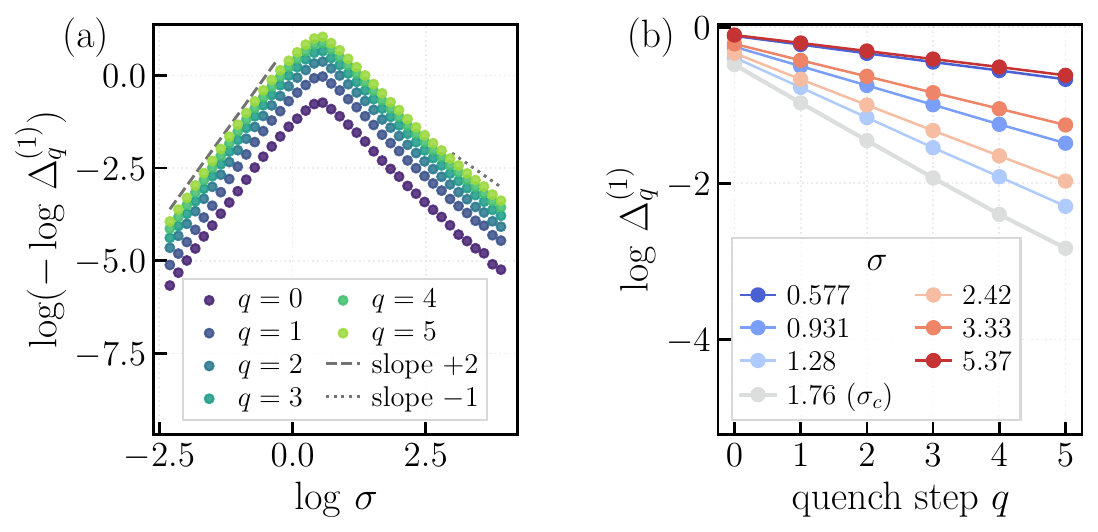} 
    \caption{Disorder dependence and multi-quench suppression of the first-moment gap for the open XXZ chain with $L=6$, $J_{xy}=1$, $J_z=0.6$, and $S^z=0$, using $5000$ disorder realizations. Each gap is sampled at the last recorded time within the corresponding evolution segment. (a) $\ln(-\ln\Delta_q^{(1)})$ versus $\ln\sigma$ for $q=0,\ldots,5$. The dashed and dotted guides have slopes $+2$ and $-1$, respectively, corresponding to the weak- and strong-disorder scalings in Eqs.~\eqref{eq:scaling_smallsigma} and \eqref{eq:scaling_largesigma}. (b) $\ln\Delta_q^{(1)}$ versus $q$ at the indicated disorder strengths. The approximately linear decrease is consistent with exponential suppression bounded by Eq.~\eqref{eq:gap_q_bound}, with the fastest suppression near $\sigma_c\simeq1.76$. Lines connect the numerical data.}
    \label{fig:gap_vs_sigma}
\end{figure}

We numerically evaluate the first-moment gap at late times within each evolution segment. Fig.~\ref{fig:gap_vs_sigma}(a) plots $\ln(-\ln\Delta_q^{(1)})$ against $\ln\sigma$ for several quench counts. The slopes approach $+2$ and $-1$ on the weak and strong-disorder sides respectively, consistent with $\ln\Delta^{(1)}\sim-\sigma^2$ and $\ln\Delta^{(1)}\sim-\sigma^{-1}$. Thus, although the perturbative expressions describe the initial decay, their leading disorder dependence empirically also captures the observed late-time plateaus. Fig.~\ref{fig:verify_xxz} suggests that the initial reduction of the gap largely sets the plateau height. The higher height means more close to Haar random. Fig.~\ref{fig:gap_vs_sigma}(b) further shows an approximately linear decrease of $\ln\Delta_q^{(1)}$ with $q$, consistent with exponential suppression of the gap. The fastest decay occurs near $\sigma_c\simeq1.76\sim J_{xy}=1$, identifying the optimal disorder strength for this finite-chain protocol.

\paragraph{\color{blue}Applications---}\label{sec:classicalshadow}
A direct application of the multi-quench protocol is classical-shadow tomography with a global-Haar measurement channel~\cite{Huang_2020}. A unitary $2$-design reproduces this channel, enabling fidelity estimation with respect to a target pure state and estimation of observable expectation values. Disordered spin chains have also been proposed as a route to efficient shadow tomography through parametrically small shadow norms~\cite{Zhou_2024}.

We illustrate this application through pure-state fidelity estimation for two physically relevant states, the cluster state $\ket{\psi_c}=\prod_j\mathrm{CZ}_{j,j+1}\ket{++\cdots+}_l$ and the GHZ state $\ket{\psi_\GHZ}=(\ket{00\cdots 0}_l+\ket{11\cdots 1}_l)/\sqrt 2$, where $l$ labels logical qubits. 
To stay inside the conserved $S^z=0$ sector of the XXZ chain, we use a dual-rail encoding $\ket{0}_l=\ket{01}$, $\ket{1}_l=\ket{10}$, so each pair of physical sites carries one logical qubit at zero magnetization. With $L=10$ physical sites and $L_l=5$ logical qubits, the encoded states $\ket{\psi_c}$ and $\ket{\psi_\GHZ}$ live in a $2^{L_l}=32$-dimensional subspace of the $\binom{L}{L/2}=252$-dimensional $S^z=0$ sector, and the global-Haar inverse channel $\mathscr{M}^{-1}(\rho)=(d+1)\rho-\id$ uses $d=252$.

The protocol prepares $\rho\approx\ket\psi\bra\psi$, applies a sampled unitary $U(t)$ generated by the disordered XXZ chain, and measures the evolved state in the computational basis, yielding an outcome $b$. Classical post-processing constructs $\rho_b=U^\dagger\ket b\bra b U$ and applies the global-Haar inverse channel, $\mathscr{M}^{-1}(\rho_b)=(d+1)\rho_b-\id$. Averaging these snapshots over sampled unitaries and measurement outcomes gives $\hat\rho$ and the fidelity estimator $\bra\psi\hat\rho\ket\psi$. This inverse channel is exact for a unitary $2$-design. Deviations of the actual ensemble from a $2$-design can introduce systematic bias. The multi-quench protocol reduces these deviations.

\begin{figure}
    \centering
    \includegraphics[width=0.8\linewidth]{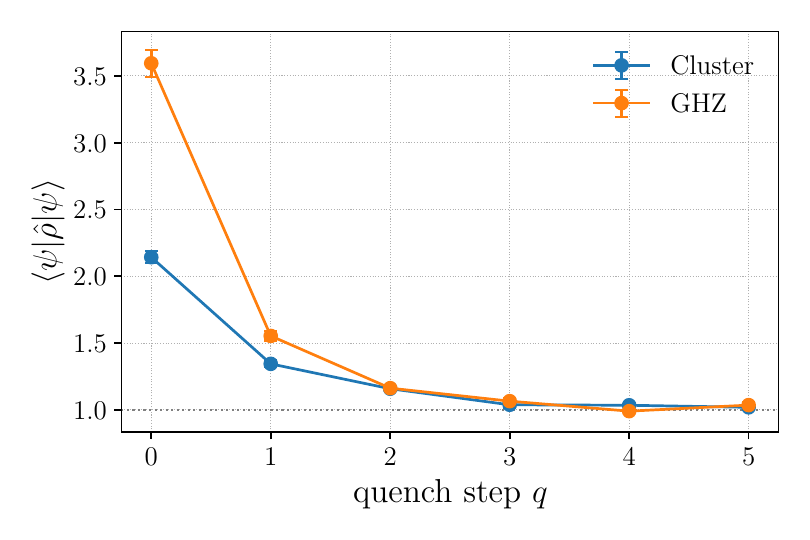}
    \caption{Classical-shadow fidelity estimation using the multi-quench XXZ ensemble. The cluster state (blue) and GHZ state (orange) are encoded in $L_l=5$ logical qubits using $L=10$ physical spins in the $S^z=0$ sector. Markers show $\bra\psi\hat\rho\ket\psi$. Error bars denote one standard error over $5000$ disorder realizations per $q$, and the horizontal dotted line marks the true fidelity of unity. The upward bias at $q=0$ is strongly reduced by additional quenches, with both estimates approaching unity by $q\simeq4$. Parameters are $J_{xy}=1$, $J_z=0.6$, $\sigma= 1.76 \simeq\sigma_c$, and total evolution time $t=30$. The $q$ switches occur at $t_a=5a$ for $a=1,\ldots,q$.}
    \label{fig:classicalshadowcomparison}
\end{figure}

Fig.~\ref{fig:classicalshadowcomparison} shows the fidelity estimator for the encoded cluster and GHZ states as a function of $q$ at $\sigma_c$. At $q=0$, the estimates are approximately $2.1$ and $3.6$, respectively, exceeding the true fidelity of unity because the single-quench ensemble does not reproduce the global-Haar measurement channel. Additional quenches strongly reduce this bias, bringing both estimates close to unity by $q\simeq4$. The standard errors also decrease with $q$ and are approximately $0.02$ for both states at $q=4$. Further quenches provide only modest improvements at the fixed sampling budget used here.

\paragraph{\color{blue}Discussion---}\label{sec:summary}

In this work, we analytically determine the optimal disorder strength for generating unitary $k$-designs from the dynamics of a disordered XXZ chain. We show that the moment gap $\Delta^{(k)}$ between the $k$-th moment operator of the dynamically generated ensemble and that of the Haar ensemble depends non-monotonically on the disorder strength $\sigma$, reaching a minimum at $\sigma_c\sim J_{xy}$, where $J_{xy}$ sets the characteristic scale of the non-random interacting dynamics. Our analysis is based on a replica/operator perturbation theory controlled by $\sigma/J_{xy}$ in the weak-disorder regime and by $J_{xy}/\sigma$ in the strong-disorder regime, and applies to arbitrary $k$ for the full interacting XXZ chain. Numerical results confirm the predicted asymptotic scalings. As an independent check, a direct perturbative calculation of the FP gap yields the same weak- and strong-disorder scalings~\cite{SM}, indicating that the non-monotonic optimum is a robust feature of disordered spin-chain dynamics rather than an artifact of the moment-operator expansion. Finally, a multi-quench protocol amplifies this single-quench optimum: $q+1$ segments with independent disorder realizations suppress the moment gap to a value no larger than $(\Delta^{(k)})^{q+1}$, enabling rapid convergence toward the Haar ensemble. We demonstrate this by generating an approximate unitary $2$-design from disordered XXZ dynamics for classical-shadow tomography.

Several directions remain open. Jointly optimizing the quench times $\{\Delta t_a\}$ together with $\sigma$, e.g., variationally, could further improve the protocol. Extending the FP calculation to $k=2$ would give a microscopic handle on the shadow-estimator variance and thus the sampling cost. The relation between $k$-design convergence and standard chaos diagnostics, such as out-of-time-order correlators and the spectral form factor~\cite{MaldacenaStanford2016PRD,ChanDeLucaChalker2018,Gharibyan:2018jrp,zhou2026measuringrenyientropyusing,zhou2026generalizedloschmidtechoinformation,liErrorResilientReversalQuantum2026}, deserves systematic study across integrable and chaotic models. Moreover, since strong disorder can induce MBL and thereby prevent thermalization, it would be interesting to investigate whether other non-thermalizing phases likewise prevent the formation of unitary $k$-designs. Finally, two-dimensional arrays and platforms with long-range interactions~\cite{Hazzard:2014bx,Lukin17NV,Yan:2013fn} modify the dispersion of $H_0$, altering both perturbative tails and thereby providing an experimental knob to tune the optimal disorder and the convergence rate of higher-dimensional approximate $k$-designs.

{\em Note added.} While completing this work, we became aware of Ref.~\cite{sur2026}, which studies the realization of unitary $k$-designs using a similar model.

{\em \color{blue} Acknowledgments---} We thank Laura Foini, Sara Murciano, Jean-Yves Desaules and Romain Vasseur for helpful discussions on related topics. T.-G. Z. was supported by the Swiss National
Science Foundation under Division II (Grant No. 200020-219400). This work has received funding through the Swiss Quantum Initiative awarded by the State Secretariat for Economic Affairs, under the grant "HoloGraph". This research is supported in part by the Fonds National Suisse de la Recherche Scientifique (Schweizerischer Nationalfonds zur Förderung der wissenschaftlichen Forschung) through the Project Grant 200021\_215300.

	\bibliography{ref}

	\clearpage
	\newpage
	\appendix
	\onecolumngrid
	\renewcommand{\appendixname}{SM}
	\setcounter{secnumdepth}{2}
	\section*{Supplemental Material}

\section{First versus second-moment gaps and in the free-fermion case}
\label{app:k1k2}

In the main-text numerics, we use the first moment. Here, we compare the first- and second-moment gaps and show that the free-fermion point $J_z=0$ cannot serve as a diagnostic for $k\geq2$.

The main idea is as follows. At $J_z=0$, the Jordan--Wigner transformation maps the disordered chain onto free fermions, so every sampled unitary acts linearly on the fermionic modes, $Uc_i^\dagger U^\dagger=\sum_a u_{ai}c_a^\dagger$ with $u\in U(L)$. Consequently, the two-replica operator $\sum_{i,j}c_i^\dagger c_j\otimes c_j^\dagger c_i$ is invariant under every $U\otimes U$, but it is not a linear combination of the identity and the SWAP operator. Its non-Haar component is thus left unchanged by the moment operator, and $\Delta^{(k)}=1$ for all $k\geq2$, independent of $\sigma$, $t$, and the number of quenches. The first moment is not affected, because the fixed-magnetization sector carries an irreducible representation of $U(L)$. For $J_z\neq0$, the Ising interaction removes this invariant.

Moreover, the second moment always inherits the deviation of the first
moment. Consider operators of the form $O\otimes\id$ with traceless $O$.
Because $U\id U^\dagger=\id$, the twofold channel acts on them as
\begin{equation}
	\mathbb{E}\!\left[U^{\otimes 2}(O\otimes\id)\,U^{\dagger\otimes 2}\right]
	= \mathbb{E}\!\left[U O U^\dagger\right]\otimes\id ,
\end{equation}
so they evolve exactly like the first-moment mode $O$. They are also
non-Haar modes at $k=2$: tracelessness of $O$ makes them orthogonal to
both Haar-invariant operators, $\id\otimes\id$ and the swap operator, so
Haar averaging maps them to zero. The second-moment deviation therefore
contains a complete copy of the first-moment deviation, and the
second moment also has additional non-Haar modes, such as $O\otimes P$
with $O$ and $P$ traceless, that can only increase it. This gives
$\Delta^{(2)}\geq\Delta^{(1)}$ for any ensemble: an ensemble cannot be
closer to a unitary $2$-design than it is to a $1$-design.

We verify this by constructing the first- and second-moment operators directly for an open chain with $L=4$ in the $S^z=0$ sector, $d=6$, so that the $k=2$ moment operator has dimension $d^4=1296$. We use the same Pauli convention as in Eq.~\eqref{eq:HXXZ} and the same parameters as the other numerical figures in the main text: $J_{xy}=1$ and Gaussian random fields. Here we average over $10^4$ disorder realizations per data point, and we compute $\Delta^{(k)}$ as the largest singular value of $\Mk{\nu}-\Mk{\mu_H}$.

\begin{figure*}[ht]
    \centering
    \includegraphics[width=0.95\textwidth]{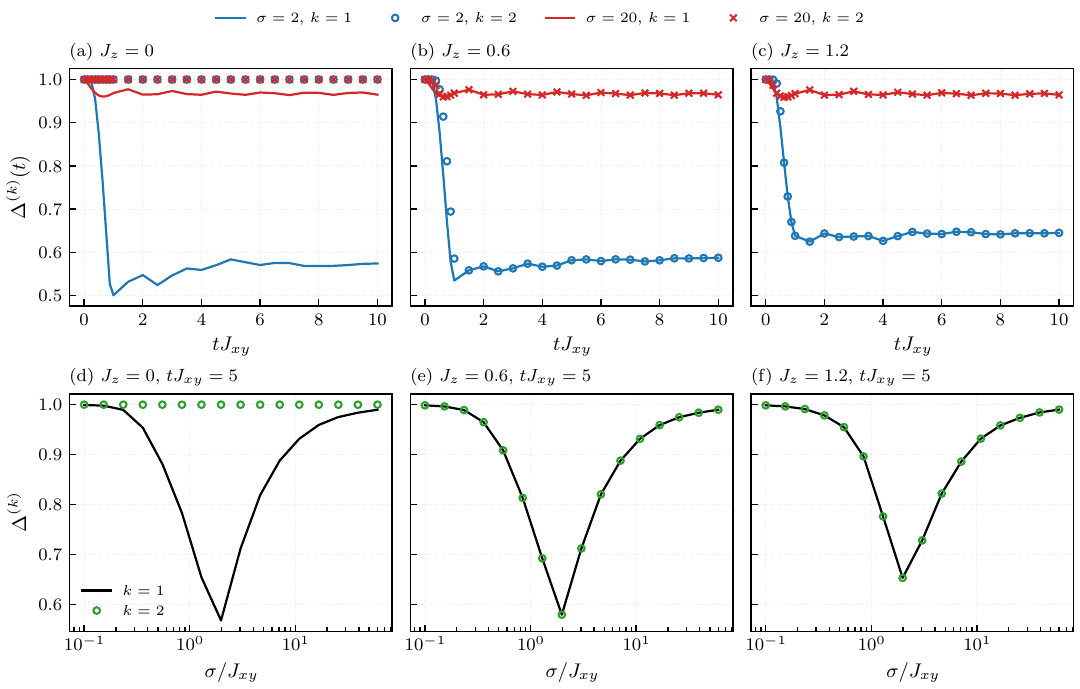}
    \caption{First and second-moment gaps of the disordered XXZ chain with $L=4$, open boundaries, $S^z=0$ ($d=6$), and $J_{xy}=1$, using $10^4$ disorder realizations per point and the same Pauli convention as the other numerical figures. Lines show $\Delta^{(1)}$ and open markers show $\Delta^{(2)}$. (a)--(c) Time dependence at $\sigma=2$ (blue) and $\sigma=20$ (red) for $J_z=0$, $0.6$, and $1.2$. (d)--(f) Disorder dependence at $tJ_{xy}=5$. At the free-fermion point $J_z=0$, $\Delta^{(2)}=1$ at all times and disorder strengths, whereas $\Delta^{(1)}$ decays. For $J_z\neq0$, the two gaps coincide after a short transient.}
    \label{fig:k1k2}
\end{figure*}

Fig.~\ref{fig:k1k2} shows the results. At $J_z=0$, $\Delta^{(2)}=1$ to machine precision at all times and disorder strengths, whereas $\Delta^{(1)}$ decays and reaches a minimum of about $0.57$ at $tJ_{xy}=5$. For $J_z=0.6$ and $1.2$, $\Delta^{(2)}$ and $\Delta^{(1)}$ differ only during a short initial transient, $tJ_{xy}\lesssim 1$ and $tJ_{xy}\lesssim 0.5$, respectively. For $tJ_{xy}\geq 1.5$, the two gaps agree to within $10^{-14}$ for both disorder strengths and across the full disorder sweep, i.e., the bound $\Delta^{(2)}\geq\Delta^{(1)}$ is saturated. Hence, for $k\geq2$ the free-fermion point $J_z=0$ fails as a diagnostic, while for $J_z\neq0$ the first-moment results of the main text are representative of the second moment, at least for this system size.

\section{Direct derivation of the frame potential in the weak- and strong-disorder regimes}
\label{app:fp_asymptotics}

In this section, we calculate the first-order frame potential of the single-quench unitary ensemble generated by the disordered XXZ chain introduced in the main text. Using perturbation theory, we treat the weak- and strong-disorder regimes separately and derive the leading nonvanishing corrections directly from the Hamiltonian.

Throughout, we set $\hbar=1$ and take $L$ to be even. All physical traces are restricted to the zero-magnetization sector $\mathcal H_n$ with $n=L/2$ and dimension $d=\binom{L}{n}$. The operators
$X_i,Y_i,Z_i$ are Pauli matrices on site $i$. We impose open boundary conditions. The disordered XXZ chain is described by
\begin{equation}
 H_{\boldsymbol h}=H_0+H_Z, \ 
 H_0=\sum_{i=1}^{L-1}
 \left[
 J_{xy}(X_iX_{i+1}+Y_iY_{i+1})
 +J_zZ_iZ_{i+1}
 \right], \ 
 H_Z=\sum_{i=1}^{L}h_iZ_i,
 \label{eq:fpapp_model}
\end{equation}
where $h_i\sim\mathcal{N}(0,\sigma^2)$ is an independent random disorder field at each site, and $\sigma$ characterizes the disorder strength. We consider the single-quench case, in which the disorder realization is fixed throughout the entire evolution time $t$.

The definition of the frame potential (FP)~\cite{Roberts_2017,Cotler_2017,Brown_2018,Gross_2007,Roy_2009} is
 \begin{equation}\label{eq:FP_def}
     F_{\nu}^{(k)}(t)= \underset{U,V\sim\nu}{\bE}
 \left| \Tr\left[U(t)V^{\dagger}(t)\right] \right|^{2k},
 \end{equation}
 where $U$ and $V$ are drawn independently from $\nu$. 

We focus on the first-order FP. The two disorder configurations $\boldsymbol h$ and
$\boldsymbol h'$ entering the first-order FP
\begin{equation}
 F^{(1)}(t,\sigma)=
 \mathbb E_{\boldsymbol h,\boldsymbol h'}
 \left|
 \operatorname{Tr}\left[
 U_{\boldsymbol h}(t)
 U_{\boldsymbol h'}^\dagger(t)
 \right]
 \right|^2
 \label{eq:fpapp_definition}
\end{equation}
are independent. Here, $U_{\boldsymbol h}(t)=e^{-iH_{\boldsymbol h}t}$, and the total evolution time $t$ itself is fixed and not averaged. The Haar value is
$F_{\mathrm {Haar}}^{(1)}=1$. We thus define the first-order FP gap as
\begin{equation}
 \Delta_{\mathrm{FP}}^{(1)}(t,\sigma)
 =F^{(1)}(t,\sigma)-1.
 \label{eq:fpapp_gap_definition}
\end{equation}
All disorder asymptotics below are taken at fixed finite $L$
and fixed $t>0$.

\subsection{Moment-operator identity and reference values}
\label{app:fp_identity}

We first introduce the averaged first-order moment operator
\begin{equation}
 M^{(1)}
 =
 \mathbb E_{\boldsymbol h}
 \left[
 U_{\boldsymbol h}\otimes U_{\boldsymbol h}^*
 \right].
\end{equation}
Independence of the two disorder samples gives
\begin{align}
 \operatorname{Tr}_{\mathrm{op}}
 \left[
 (M^{(1)})^\dagger M^{(1)}
 \right]
 &=
 \mathbb E_{\boldsymbol h,\boldsymbol h'}
 \operatorname{Tr}_{\mathrm{op}}
 \left[
 (U_{\boldsymbol h}^\dagger U_{\boldsymbol h'})
 \otimes
 (U_{\boldsymbol h}^T U_{\boldsymbol h'}^*)
 \right]
 \nonumber\\
 &=
 \mathbb E_{\boldsymbol h,\boldsymbol h'}
 \left|
 \operatorname{Tr}
 (U_{\boldsymbol h}^\dagger U_{\boldsymbol h'})
 \right|^2
 =
 F^{(1)}.
 \label{eq:fpapp_moment_identity}
\end{align}
Here, $\operatorname{Tr}_{\mathrm{op}}$ denotes a trace on the
doubled space.

With unnormalized vectorization
$|I_d\rangle\!\rangle=\sum_x|x,x\rangle$, the Haar moment is $M_{\mathrm H}^{(1)}
 =
 \frac{
 |I_d\rangle\!\rangle
 \langle\!\langle I_d|
 }{d}$.

Since
$M^{(1)}M_{\mathrm H}^{(1)}
=M_{\mathrm H}^{(1)}M^{(1)}
=M_{\mathrm H}^{(1)}$,
Eq.~\eqref{eq:fpapp_moment_identity} implies
\begin{equation}
 \Delta_{\mathrm{FP}}^{(1)}
 =
 \left\|
 M^{(1)}-M_{\mathrm H}^{(1)}
 \right\|_2^2.
 \label{eq:fpapp_haar_identity}
\end{equation}
This is the usual frame-potential identity, with
$\|\cdot\|_2$ the Hilbert--Schmidt norm.

At $\sigma=0$, every sample produces the same unitary, so
$F^{(1)}(t,0)=d^2$. In contrast, strong local disorder suppresses
coherences between distinct computational-basis states, and its
atomic-limit reference is $F^{(1)}=d$, as we will derive below.

Consequently, the appropriate plateau-normalized
Haar-subtracted quantities are
\begin{equation}
 R_{\mathrm w}(t,\sigma)
 =
 \frac{F^{(1)}(t,\sigma)-1}{d^2-1},
 \qquad
 R_{\mathrm s}(t,\sigma)
 =
 \frac{F^{(1)}(t,\sigma)-1}{d-1}.
\label{eq:fpapp_normalizations}
\end{equation}
The two denominators refer to different disorder limits of the
same frame-potential difference. Here, $R_{\mathrm w}$ and $R_{\mathrm s}$ denote the normalizations appropriate to the weak- and strong-disorder limits, respectively.

\subsection{Weak-disorder expansion}
\label{app:fp_weak}

In the interaction picture of $H_0$, the time-evolution operator is
\begin{equation}
 U_{\boldsymbol h}(t)
 =
 e^{-iH_0t}W_{\boldsymbol h}(t),
 \qquad
 W_{\boldsymbol h}(t)
 =
 \mathcal T
 \exp\left[
 -i\sum_i h_i\int_0^t Z_i(s)\,ds
 \right].
\end{equation}
Here, the time-evolved operator is defined as $ Z_i(t)=e^{iH_0t}Z_ie^{-iH_0t}.$ Cyclicity of the trace removes $e^{-iH_0 t}$ from
Eq.~\eqref{eq:fpapp_definition}:
\begin{equation}
 F^{(1)}
 =
 \mathbb E_{\boldsymbol h,\boldsymbol h'}
 \left|
 \operatorname{Tr}
 \left(
 W_{\boldsymbol h}
 W_{\boldsymbol h'}^\dagger
 \right)
 \right|^2.
 \label{eq:fpapp_weak_IP}
\end{equation}
No approximation in $J_z$ is involved.

Define
\begin{equation}
 A_i(t)=\int_0^t Z_i(s)\,ds,
 \qquad
 B_{ij}(t)=
 \int_0^tds\int_0^t ds'\,
 Z_i(s)Z_j(s').
 \label{A_definition_weak_disorder}
\end{equation}
Then
\begin{equation}
 W_{\boldsymbol h}
 =
 I_d
 -i\sum_i h_iA_i
 -\sum_{ij}h_ih_jB_{ij}
 +O(\boldsymbol h^3).
\end{equation}
Using the definitions in Eq.~\eqref{A_definition_weak_disorder}, the Hermitian operators $A_i$ obey
\begin{equation}
 B_{ii}+B_{ii}^\dagger=A_i^2.
 \label{eq:fpapp_B_identity}
\end{equation}
Also, in the half-filled sector $ \operatorname{Tr}Z_i
 =
 \operatorname{Tr}A_i
 =
 0$. Writing
$t_{\boldsymbol h,\boldsymbol h'}
=\operatorname{Tr}
(W_{\boldsymbol h}W_{\boldsymbol h'}^\dagger)$,
we obtain
\begin{align}
 t_{\boldsymbol h,\boldsymbol h'}
 =
 d
 -\sum_{ij}h_ih_j\operatorname{Tr}B_{ij}
 -\sum_{ij}h_i'h_j'
 \operatorname{Tr}B_{ij}^\dagger
 +\sum_{ij}h_ih_j'
 \operatorname{Tr}(A_iA_j)
 +O((\boldsymbol h,\boldsymbol h')^3).
 \label{eq:fpapp_trace_Dyson}
\end{align}
There is no linear term because each $A_i$ is traceless. Using the Gaussian covariances $\mathbb E [h_ih_j]=
 \mathbb E [h_i'h_j']=
 \sigma^2\delta_{ij}, \ 
 \mathbb E [h_ih_j']=0$, together with Eq.~\eqref{eq:fpapp_B_identity}, we obtain
\begin{equation}
 \label{eq:fpapp_weak_result}
 F^{(1)}(t,\sigma)= \mathbb E_{\boldsymbol h,\boldsymbol h'}
 [\left|
 t_{\boldsymbol h,\boldsymbol h'}
 \right|^2]
 =
 d^2
 -2d\sigma^2\sum_i
 \operatorname{Tr}[A_i(t)^2]
 +O(\sigma^4).
\end{equation}
Odd powers vanish under the centered Gaussian average. The remainder is evaluated at fixed $t$ and $L$.

Define the nonnegative quantity
\begin{equation}
 \mathcal W(t)
 =
 \frac1d\sum_i
 \operatorname{Tr}[A_i(t)^2]
 \geq0,
 \qquad
 C_i(\tau)
 =
 \frac1d
 \operatorname{Tr}[Z_i(\tau)Z_i].
\end{equation}
Time-translation invariance of the sector trace gives
\begin{align}
 \mathcal W(t)
 =\sum_i\int_0^t ds\int_0^t ds'\,
 C_i(s-s')
 =
 2\sum_i\int_0^t
 (t-\tau)
 \operatorname{Re}C_i(\tau)\,d\tau.
 \label{eq:fpapp_weak_correlator}
\end{align}

Equivalently, let $\{|a\rangle\}$ denote the eigenbasis of $H_0$, with $H_0|a\rangle=E_a|a\rangle$. Then
\begin{equation}
 \mathcal W(t)
 =
 \frac1d\sum_{i=1}^L \sum_{a,b=1}^d
 |\langle a|Z_i|b\rangle|^2
 f_t(E_a-E_b),
 \label{eq:fpapp_weak_spectral}
\end{equation}
where
\begin{equation}
 f_t(\omega)=
 \begin{cases}
 \displaystyle
 \frac{4\sin^2(\omega t/2)}{\omega^2},
 & \omega\neq0,
 \\[6pt]
 t^2,
 & \omega=0.
 \end{cases}
\end{equation}
This expression includes degeneracies and applies to the
full interacting XXZ Hamiltonian. 

We now extract the disorder dependence of the FP. When the perturbative correction is small,
\begin{align}
 \ln\frac{F^{(1)}(t,\sigma)}{d^2}
 =
 -2\sigma^2\mathcal W(t)
 +O(\sigma^4),
 \label{eq:fpapp_weak_logF}
\end{align}
For the plateau-normalized, Haar-subtracted quantity defined in Eq.~\eqref{eq:fpapp_normalizations}, this gives 
\begin{align}
 \ln R_{\mathrm w}(t,\sigma)
 =
 -\frac{2d^2}{d^2-1}
 \sigma^2\mathcal W(t)
 +O(\sigma^4).
 \label{eq:fpapp_weak_loggap}
\end{align}

Thus, the leading weak-disorder correction to the normalized
logarithm is proportional to $-\sigma^2$, provided
$\mathcal W(t)>0$. The unnormalized logarithm instead
contains the offset $\ln(d^2-1)$.

As a short-time check,
$A_i(t)=tZ_i+O(t^2)$ and
$\mathcal W(t)=Lt^2+O(t^4)$, so $ F^{(1)}(t,\sigma)
 =
 d^2\left[
 1-2L\sigma^2t^2+\cdots
 \right]$. In general, this does not imply a departure from the plateau value $d^2$ with a coefficient linear in $t$, which would require additional
assumptions on the decay and integrability of the clean
correlations.

\subsection{Strong-disorder expansion}
\label{app:fp_atomic}

In the strong-disorder regime, we decompose the Hamiltonian into parts that are diagonal and off-diagonal in the $Z$ basis:
\begin{equation}
 H_{\boldsymbol h}
 =
 D_{\boldsymbol h}+V,
 \qquad
 D_{\boldsymbol h}
 =
 \sum_{i=1}^L h_iZ_i
 +J_z\sum_{i=1}^{L-1}Z_iZ_{i+1},
 \qquad
 V=
 J_{xy}\sum_{i=1}^{L-1}
 (X_iX_{i+1}+Y_iY_{i+1}).
 \label{eq:fpapp_strong_split}
\end{equation}
The diagonal Ising interaction is retained in $D_{\boldsymbol h}$ as part of the zeroth-order Hamiltonian, while $V$, which contains the $XX$ and $YY$ terms of the XXZ chain, is treated as a perturbation.

For a computational-basis configuration
$\mathbf{s}=(s_1,\ldots,s_L)$ in the fixed-magnetization sector,
we use the $Z$-basis $|\mathbf{s}\rangle
=
|s_1\rangle\otimes\cdots\otimes|s_L\rangle$, $s_i\in\{\uparrow,\downarrow\}$, such that $Z_i|\mathbf{s}\rangle
    =
    z_i(\mathbf{s})|\mathbf{s}\rangle, \ 
    z_i(\mathbf{s})=\pm1$. Here, $\mathbf{s}$ is simply a label for a many-body spin configuration. Since $D_{\boldsymbol h}$ is diagonal in this basis, we have $ D_{\boldsymbol h}|\mathbf{s}\rangle
    =
    E_{\mathbf{s}}(\boldsymbol h)|\mathbf{s}\rangle$,
where $E_{\mathbf{s}}(\boldsymbol h)
    =
    \sum_{i=1}^L h_i z_i(\mathbf{s})
    +
    E_z(\mathbf{s}), \ 
    E_z(\mathbf{s})
    =
    J_z\sum_{i=1}^{L-1}
    z_i(\mathbf{s})z_{i+1}(\mathbf{s})$.

We first evaluate the FP exactly in the atomic limit, $J_{xy}=0$. In this limit, the time evolution is diagonal, $U_{\boldsymbol h}(t)|\mathbf{s}\rangle
    =
    e^{-iE_{\mathbf{s}}(\boldsymbol h)t}
    |\mathbf{s}\rangle$.
Therefore,
\begin{align}
    F_{\mathrm{at}}^{(1)}(t,\sigma)
    &=
    \mathbb E_{\boldsymbol h,\boldsymbol h'}
    \left|
    \sum_{\mathbf{s}}
    e^{-i[
    E_{\mathbf{s}}(\boldsymbol h)
    -
    E_{\mathbf{s}}(\boldsymbol h')
    ]t}
    \right|^2
    =
    \sum_{\mathbf{s},\mathbf{s}'}
    \mathbb E_{\boldsymbol h,\boldsymbol h'}
    \exp\left[
    -it
    \sum_i(h_i-h_i')
    \left(
    z_i(\mathbf{s})
    -
    z_i(\mathbf{s}')
    \right)
    \right].
\end{align}
The deterministic Ising contribution $E_z(\mathbf{s})$
cancels between the two independent disorder realizations.
Since $h_i-h_i'\sim\mathcal N(0,2\sigma^2)$, the Gaussian
average gives
\begin{equation}
    F_{\mathrm{at}}^{(1)}(t,\sigma)
    =
    \sum_{\mathbf{s},\mathbf{s}'}
    \exp\left[
    -\sigma^2t^2
    \sum_i
    \left(
    z_i(\mathbf{s})-z_i(\mathbf{s}')
    \right)^2
    \right].
\end{equation}

For any fixed computational-basis configuration
$\mathbf{s}$ in the $n$-particle sector, configurations $\mathbf{s}'$ at Hamming distance $2r$ are obtained by selecting $r$ occupied and $r$ unoccupied sites and exchanging their occupations. Hence
\begin{equation}
 F_{\mathrm{at}}^{(1)}(t,\sigma)
 =
 d\sum_{r=0}^{\min(n,L-n)}
 \binom{n}{r}
 \binom{L-n}{r}
 e^{-8r\sigma^2t^2}.
 \label{eq:fpapp_atomic_exact}
\end{equation}
Here, the prefactor $d$ arises because there are $d$ possible choices of $\mathbf{s}$.
In particular,
\begin{equation}
 F_{\mathrm{at}}^{(1)}(t,\sigma)
 =
 d+O_L(e^{-8\sigma^2t^2}),
 \qquad
 \sigma t\gg1.
 \label{eq:fpapp_atomic_plateau}
\end{equation}
Thus, the atomic strong-disorder reference value of the Haar-subtracted FP is $d-1$.

We then calculate the leading two-hop correction.
The following calculation retains the first nonvanishing order in the hopping term $V$, which is second order in $J_{xy}$.

The hopping operator connects two computational-basis
configurations $\mathbf{s}$ and $\mathbf{s}'$ that differ by a single nearest-neighbor exchange,
$01\leftrightarrow10$. Its matrix element is
\begin{equation}
 g =
 \langle 10|
 J_{xy}(X_iX_{i+1}+Y_iY_{i+1})
 |01\rangle =
 2J_{xy}.
 \label{eq:fpapp_hopping_convention}
\end{equation}

For a connected pair
$\mathbf{s}\leftrightarrow\mathbf{s}'$, define the diagonal-energy difference $\delta_{\mathbf{s}'\mathbf{s}}
 =
 E_{\mathbf{s}'}(\boldsymbol h)
 -
 E_{\mathbf{s}}(\boldsymbol h)$. The first-order transition amplitude in the interaction picture
of $D_{\boldsymbol h}$ is
\begin{equation}
 \langle \mathbf{s}'|
 U_{\boldsymbol h}(t)
 |\mathbf{s}\rangle
 =
 -ig\,
 e^{-iE_{\mathbf{s}'}(\boldsymbol h)t}
 \int_0^t
 e^{i\delta_{\mathbf{s}'\mathbf{s}}s}\,ds
 +\cdots.
 \label{eq:fpapp_transition_amplitude}
\end{equation}
Here, the interaction-picture decomposition is
\begin{equation}
 U_{\boldsymbol h}(t)
 =
 e^{-iD_{\boldsymbol h}t}W(t),
 \qquad
 W(t)
 =
 \mathcal T
 \exp\left[
 -i J_{xy} \sum_i \int_0^t (X_iX_{i+1}+Y_iY_{i+1})(s)\,ds
 \right].
\end{equation}
The corresponding transition probability, to second order in $g$, is therefore
\begin{equation}
    p_{\mathbf{s}'\leftarrow\mathbf{s}}^{[2]}(t)
 =
 g^2
 \left|
 \int_0^t
 e^{i\delta_{\mathbf{s}'\mathbf{s}}s}\,ds
 \right|^2
 =
 2g^2
 \int_0^t
 (t-\tau)
 \cos\!\left(
 \delta_{\mathbf{s}'\mathbf{s}}\tau
 \right)
 d\tau .
\label{eq:fpapp_transition_probability}
\end{equation}
Here, the superscript $[2]$ denotes second order in the hopping
perturbation and should not be confused with the order of the
FP.

To average over the random fields, suppose that $\mathbf{s}$ and $\mathbf{s}'$ differ by an exchange on the physical bond $(i,i+1)$. Since the two spins on this bond have opposite $Z$ eigenvalues before the exchange, the full diagonal-energy difference is $ \delta_{\mathbf{s}'\mathbf{s}}
 =
 \pm2(h_i-h_{i+1})
 +
 \delta_z(\mathbf{s}',\mathbf{s})$, where $\delta_z(\mathbf{s}',\mathbf{s})
 =
 E_z(\mathbf{s}')
 -
 E_z(\mathbf{s})$.
For the nearest-neighbor Ising interaction, $|\delta_z(\mathbf{s}',\mathbf{s})|
 \leq 4|J_z|$. Because $h_i$ and $h_{i+1}$ are independent Gaussian variables
with variance $\sigma^2$, the random part
$\pm2(h_i-h_{i+1})$ has variance $8\sigma^2$. Therefore, $\mathbb E_{\boldsymbol h}
 \left[
 e^{i\delta_{\mathbf{s}'\mathbf{s}}\tau}
 \right]
 =
 e^{-4\sigma^2\tau^2}
 e^{i\delta_z(\mathbf{s}',\mathbf{s})\tau}$.

Using the relation above and averaging over the random fields, we obtain
\begin{equation}
 \overline{
 p_{\mathbf{s}'\leftarrow\mathbf{s}}^{[2]}(t)
 }
 =
 2g^2
 C_\sigma
 \!\left(
 t;
 \delta_z(\mathbf{s}',\mathbf{s})
 \right),
 \label{eq:fpapp_probability_kernel}
\end{equation}
where
\begin{equation}
 C_\sigma(t;\delta_z)
 =
 \int_0^t
 (t-\tau)
 e^{-4\sigma^2\tau^2}
 \cos(\delta_z\tau)\,d\tau.
 \label{eq:fpapp_gaussian_kernel}
\end{equation}

When $\delta_z=0$, this integral can be evaluated explicitly:
\begin{equation}
 C_\sigma(t;0)
 =
 \frac{\sqrt{\pi}t}{4\sigma}
 \operatorname{erf}(2\sigma t)
 -
 \frac{1-e^{-4\sigma^2t^2}}{8\sigma^2}.
 \label{eq:fpapp_kernel_exact}
\end{equation}
In the limit $\sigma t\gg1$ and $\sigma\gg |J_z|$,
the Gaussian kernel is concentrated around
$\tau\sim\sigma^{-1}$, so that
$\cos(\delta_z\tau)=1+O(\delta_z^2/\sigma^2)$. Hence
\begin{equation}
 C_\sigma(t;\delta_z)
 =
 \frac{\sqrt{\pi}t}{4\sigma}
 +
 O\!\left(
 \frac{1}{\sigma^2}
 +
 \frac{\delta_z^2t}{\sigma^3}
 \right).
 \label{eq:fpapp_kernel_asymptotic}
\end{equation}
Consequently,
\begin{equation}
 \overline{
 p_{\mathbf{s}'\leftarrow\mathbf{s}}^{[2]}(t)
 }
 =
 \frac{\sqrt{\pi}g^2t}{2\sigma}
 +
 O\!\left(
 \frac{g^2}{\sigma^2}
 +
 \frac{g^2J_z^2t}{\sigma^3}
 \right)
 =
 \frac{2\sqrt{\pi}J_{xy}^2t}{\sigma}
 +\cdots .
\label{eq:fpapp_probability_asymptotic}
\end{equation}

The factor $1/\sigma$ originates from the area of the Gaussian memory kernel $ \int_0^\infty
 e^{-4\sigma^2\tau^2}\,d\tau
 =
 \frac{\sqrt{\pi}}{4\sigma}$. A finite Ising detuning affects only subleading terms in the large-$\sigma$ expansion. 

We next relate the return probabilities to the FP by introducing the projector onto the diagonal operator subspace in the computational $Z$-basis,
\begin{equation}
 P_{\mathrm d}
 =
 \sum_{\mathbf{s}}
 |\mathbf{s},\mathbf{s}\rangle
 \langle\mathbf{s},\mathbf{s}|.
 \label{eq:fpapp_Pd}
\end{equation}
Here
$|\mathbf{s},\mathbf{s}\rangle
\equiv
|\mathbf{s}\rangle\otimes|\mathbf{s}\rangle$
is a basis state in the doubled Hilbert space.
The projector $P_{\mathrm d}$ has rank $d$.

Writing $E=M^{(1)}-P_{\mathrm d}$, so that $M^{(1)}=P_{\mathrm d}+E$, we express the first-order FP exactly as
\begin{equation}
    F^{(1)}=
 \operatorname{Tr}_{\mathrm{op}}
 \left[
 (M^{(1)})^\dagger M^{(1)}
 \right]
 =
 d
 +
 2\operatorname{Re}
 \operatorname{Tr}_{\mathrm{op}}
 (P_{\mathrm d}E)
 +
 \|E\|_2^2.
 \label{eq:fpapp_FP_expand_Pd}
\end{equation}
For each computational-basis configuration $\mathbf{s}$, $\langle
 \mathbf{s},\mathbf{s}
 |
 M^{(1)}
 |
 \mathbf{s},\mathbf{s}
 \rangle
 =
 \mathbb E_{\boldsymbol h}
 \left|
 \langle\mathbf{s}|
 U_{\boldsymbol h}
 |\mathbf{s}\rangle
 \right|^2
 =
 \mathbb E_{\boldsymbol h}
 |U_{\mathbf{s}\mathbf{s}}|^2$.
Unitarity gives $|U_{\mathbf{s}\mathbf{s}}|^2
 =
 1-
 \sum_{\mathbf{s}'\neq\mathbf{s}}
 |U_{\mathbf{s}'\mathbf{s}}|^2$, and thus $\langle
 \mathbf{s},\mathbf{s}
 |
 M^{(1)}
 |
 \mathbf{s},\mathbf{s}
 \rangle
 =
 1-
 \sum_{\mathbf{s}'\neq\mathbf{s}}
 \mathbb E_{\boldsymbol h}
 |U_{\mathbf{s}'\mathbf{s}}|^2$. Substituting this result into Eq.~\eqref{eq:fpapp_FP_expand_Pd}, we obtain the exact identity
\begin{equation}
 F^{(1)}
 =
 d
 -
 2
 \sum_{\mathbf{s}}
 \sum_{\mathbf{s}'\neq\mathbf{s}}
 \mathbb E_{\boldsymbol h}
 |U_{\mathbf{s}'\mathbf{s}}|^2
 +
 \left\|
 M^{(1)}-P_{\mathrm d}
 \right\|_2^2.
 \label{eq:fpapp_exact_return_identity}
\end{equation}

In the strong-disorder hopping expansion, the atomic
contribution outside $P_{\mathrm d}$ is exponentially small. The unsuppressed terms linear in hopping connect diagonal and off-diagonal operator states. For a connected pair $\mathbf{s}\leftrightarrow\mathbf{s}'$, for example,
\begin{equation}
 \left\langle
 \mathbf{s}',\mathbf{s}
 \left|
 M^{[1]}
 \right|
 \mathbf{s},\mathbf{s}
 \right\rangle
 =
 -ig
 \int_0^t
 e^{-4\sigma^2\tau^2
 -i\delta_z(\mathbf{s}',\mathbf{s})\tau}
 d\tau ,
 \label{eq:fpapp_Mfirst}
\end{equation}
which satisfies $ \left|
 \left\langle
 \mathbf{s}',\mathbf{s}
 \left|
 M^{[1]}
 \right|
 \mathbf{s},\mathbf{s}
 \right\rangle
 \right|
 \leq
 \frac{\sqrt{\pi}|g|}{4\sigma}$. Here, $M^{[1]}$ denotes the contribution linear in $g$, not the
first-order moment operator.

At fixed finite $L$, the squared Hilbert--Schmidt norm of these
terms is therefore $O_L(g^2/\sigma^2)$. The leading
$g^2t/\sigma$ contribution to the FP instead comes
from the loss of diagonal return probability in
Eq.~\eqref{eq:fpapp_exact_return_identity}.

Keeping only the leading second-order hopping contribution thus gives
\begin{equation}
 F^{(1)}(t,\sigma)
 \simeq
 d
 -
 2
 \sum_{\mathbf{s}}
 \sum_{
 \mathbf{s}':\,
 \mathbf{s}\leftrightarrow\mathbf{s}'
 }
 \overline{
 p_{\mathbf{s}'\leftarrow\mathbf{s}}^{[2]}(t)
 }.
 \label{eq:fpapp_leading_return}
\end{equation}
The omitted terms include
$O_L(J_{xy}^2/\sigma^2)$,
exponentially small atomic corrections, and fourth and higher orders in the hopping perturbation.

We now count the allowed configuration-space edges. For each physical bond $(i,i+1)$, a configuration
$\mathbf{s}$ can undergo a hopping process only if the two sites have opposite occupations, namely $01$ or $10$. An undirected configuration-space edge is specified by fixing, for example, $01$ on this bond and placing the remaining $n-1$ occupied sites among the other $L-2$ positions. Therefore, the total number of undirected edges is $N_{\mathrm e}
 =
 (L-1)
 \binom{L-2}{n-1}$.

Using $ d=\binom Ln$, one finds $ \frac{N_{\mathrm e}}{d}
 =
 \frac{n(L-n)}{L}$. At half-filling case considered here, $n=L/2$, this becomes $\frac{N_{\mathrm e}}{d}
 =
 \frac{L}{4}$. Each undirected edge corresponds to two directed transitions,
$\mathbf{s}\rightarrow\mathbf{s}'$ and
$\mathbf{s}'\rightarrow\mathbf{s}$. Hence, $\sum_{\mathbf{s}}
 \sum_{
 \mathbf{s}':\,
 \mathbf{s}\leftrightarrow\mathbf{s}'
 }
 1
 =
 2N_{\mathrm e}$.

Substituting
Eq.~\eqref{eq:fpapp_probability_asymptotic} into
Eq.~\eqref{eq:fpapp_leading_return} yields
\begin{equation}
 F^{(1)}(t,\sigma)
 \simeq
 d
 -
 \frac{
 8\sqrt{\pi}
 N_{\mathrm e}
 J_{xy}^2t
 }{\sigma}
 =
 d
 \left(
 1-
 \frac{
 2\sqrt{\pi}
 LJ_{xy}^2t
 }{\sigma}
 \right),
 \qquad n=\frac{L}{2}.
 \label{eq:fpapp_strong_result}
\end{equation}

Here and below, $\simeq$ denotes the leading term in the
large-$\sigma$ expansion while simultaneously retaining only
second order in the hopping perturbation. A conservative sufficient regime for this bare perturbative calculation at fixed finite size is $ \sigma t\gg1, \ 
 \|V\|_\infty t\ll1$. The first inequality ensures strong disorder-induced dephasing, so that the first-order moment operator is close to the projector onto the Z-basis diagonal subspace defined in Eq.~\eqref{eq:fpapp_Pd}. Consequently, in Eq.~\eqref{eq:fpapp_FP_expand_Pd}, the last term can be safely neglected. The second inequality ensures that the perturbative expansion in the hopping term remains controlled. The logarithmic expansion used below additionally requires the fractional correction to the FP to be small.

We can now extract the $1/\sigma$ scaling of the
frame-potential gap. Equation~\eqref{eq:fpapp_strong_result} immediately gives
\begin{equation}
 \ln\frac{F^{(1)}(t,\sigma)}{d}
 \simeq
 -\frac{
 2\sqrt{\pi}
 LJ_{xy}^2t
 }{\sigma}.
 \label{eq:fpapp_strong_logF}
\end{equation}

The Haar value of the first-order FP is
$F_{\mathrm H}^{(1)}=1$, so the FP gap is
\begin{equation}
 \Delta_{\mathrm{FP}}^{(1)}
 =
 F^{(1)}-1.
\end{equation}
Since its strong-disorder atomic-limit value is $d-1$, a natural plateau-normalized gap is
\begin{equation}
 R_{\mathrm s}(t,\sigma)
 =
 \frac{
 F^{(1)}(t,\sigma)-1
 }{d-1}.
\end{equation}
We use this definition rather than the simpler choice $F^{(1)}(t,\sigma)-d$, which can become negative. In contrast, since
$F^{(1)}(t,\sigma)\geq F^{(1)}_{\mathrm H}=1$, the normalized gap $R_{\mathrm s}(t,\sigma)$ is always nonnegative.

Using Eq.~\eqref{eq:fpapp_strong_result}, we have
\begin{equation}
 \ln R_{\mathrm s}(t,\sigma)
 =
 \ln
 \frac{
 F^{(1)}(t,\sigma)-1
 }{d-1}
 \simeq
 -
 \frac{
 2\sqrt{\pi}d
 }{d-1}
 \frac{
 LJ_{xy}^2t
 }{\sigma}.
 \label{eq:fpapp_strong_loggap}
\end{equation}
Thus, the deviation of the strong-disorder frame-potential gap from its atomic plateau obeys the same inverse-disorder scaling as the moment-operator gap, $-\ln R_{\mathrm s}
 \propto
 \sigma^{-1}$. Equivalently, without normalizing by the atomic plateau,
\begin{equation}
 \ln\Delta_{\mathrm{FP}}^{(1)}
 \simeq
 \ln(d-1)
 -
 \frac{
 2\sqrt{\pi}d
 }{d-1}
 \frac{
 LJ_{xy}^2t
 }{\sigma}.
 \label{eq:fpapp_strong_offset}
\end{equation}
Therefore, the raw logarithm
$\ln\Delta_{\mathrm{FP}}^{(1)}$ approaches the nonzero constant
$\ln(d-1)$ as $\sigma\rightarrow\infty$. It is the correction
to this plateau that exhibits the $1/\sigma$ behavior.

\section{Replica notation and Haar moment operator basis}\label{app:haar_basis}

	The Haar moment operator can be written in terms of the orthogonal basis
	\begin{equation}
		\Mk{\mu_H} = \sum_{\varpi,\varsigma\in S_k} \mathrm{Wg}(\varpi^{-1}\varsigma,d)\, \kket{V_d(\varpi)}\bbra{V_d(\varsigma)},
	\end{equation}
	where $S_k$ is the $k$-th symmetric group, $\mathrm{Wg}(\varpi,d)$ are the Weingarten coefficients, and $V_d(\varpi)\in\linOpCdk{}$ acts as
	\begin{equation*}
		V_d(\varpi) \ket{\psi_1}\cdots\ket{\psi_k} = \ket{\psi_{\varpi^{-1}(1)}}\cdots\ket{\psi_{\varpi^{-1}(k)}}.
	\end{equation*}
	The operator state for $O\in\linOpCdk{}$ is $\kket{O}\equiv O\otimes\idk{}\kket{\Phi^+}^{\otimes k}$ where $\kket{\Phi^+}=\sum_j\ket{j}\ket{j}/\sqrt{d}$ is the EPR pair connecting the forward $U$ and backward $U^*$ contours. The orthogonal basis becomes
	\begin{equation}\label{eq:Vd_basis}
		\kket{V_d(\varpi)} = \prod_{l=1}^{k} \sum_{j_l=1}^{d} \ket{j_{\varpi^{-1}(l)}}\ket{j_l}.
	\end{equation}

\begin{figure}[h]
		\centering
		\includegraphics[width=0.4\linewidth]{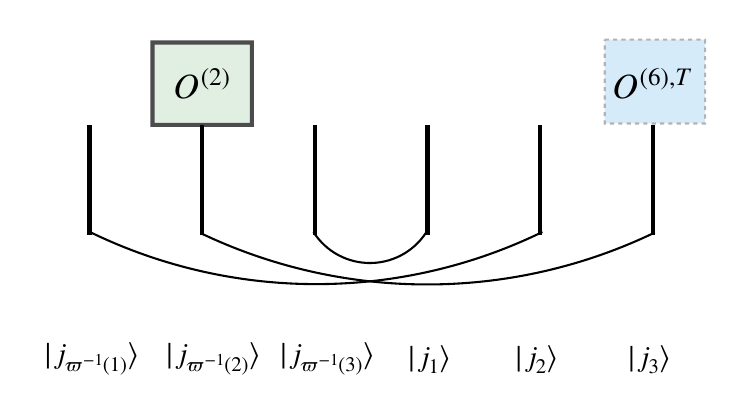}
		\caption{Graphical representation of the replica operator identity in Eq.~\eqref{eq:Op_on_Vdbasis} for $k=3$ and the permutation $\varpi^{-1}(1)=2$, $\varpi^{-1}(2)=3$, $\varpi^{-1}(3)=1$. The connecting lines pair forward and backward replicas in $\kket{V_d(\varpi)}$. Acting with $O$ on forward replica $2$ is equivalent to acting with its transpose on backward replica $6$, giving $O^{(2)}\kket{V_d(\varpi)}=O^{(6),T}\kket{V_d(\varpi)}$.}
		\label{fig:fig1replica}
	\end{figure}

	A useful identity allows one to move single-replica operators across the EPR pair: $O\otimes\idk{}\kket{\Phi^+}=\idk{}\otimes O^T\kket{\Phi^+}$. In $k$ copies, with $O^{[\alpha]}=\id\otimes\cdots\otimes O\otimes\cdots\otimes\id$ acting on the $\alpha$-th replica,
	\begin{equation}\label{eq:Op_on_Vdbasis}
		O^{(\alpha)}\kket{V_d(\varpi)} = O^{(\varpi^{-1}(\alpha)+k),T}\kket{V_d(\varpi)},
	\end{equation}
	where $O^{(\alpha)}\equiv O^{[\alpha]}\otimes\idk{}$ and $O^{(\alpha+k),T}\equiv\idk{}\otimes O^{[\alpha],T}$. Summing over replicas,
	\begin{equation}\label{eq:Opsum_idendity_Vd}
		\sum_{\alpha=1}^{k} O^{(\alpha)}\kket{V_d(\varpi)} = \sum_{\alpha=1}^{k} O^{(\alpha+k),T}\kket{V_d(\varpi)}.
	\end{equation}
	The graphical content of Eq.~\eqref{eq:Op_on_Vdbasis} is illustrated in Fig.~\ref{fig:fig1replica}.

\section{Mathematical lemmas for the moment operator}
\label{app:math_lemma}

In this section, we provide proofs of two mathematical identities used in the main text.

\paragraph{Exponential suppression of the moment gap under multiple quenches}

In the multiple-quench protocol discussed in the main text, we use the fact that the moment gap decreases exponentially with the number of quenches, $ \Delta_q^{(k)}
    \approx
    \left(\Delta^{(k)}\right)^{q+1}$. Here, we provide the argument underlying this relation.

For any matrix $A$, the spectral norm is submultiplicative,
\begin{equation}
    \|A^{q+1}\|_\infty
    \leq
    \|A\|_\infty^{q+1}.
\end{equation}
If $A$ is positive-semidefinite and Hermitian, the bound is saturated:
\begin{equation}
    \|A^{q+1}\|_\infty
    =
    \|A\|_\infty^{q+1}.
\end{equation}
In the weak-disorder regime, we apply this result to
$A=\Mk{I,\nu}$. Owing to the symmetric Gaussian disorder distribution,
$\Mk{I,\nu}$ is Hermitian and positive semidefinite. Therefore, for
$q+1$ statistically independent quenches,
\begin{equation}
    \Delta_q^{(k)}
    =
    \left(\Delta^{(k)}\right)^{q+1}.
\end{equation}
Thus, whenever $\Delta^{(k)}<1$, the moment gap is exponentially
suppressed with the number of quenches, as stated in the main text.

In the strong-disorder regime, the same reasoning applies within the
slow subspace. In this case, the relevant contraction operator is
$\idk{2}-\calK^{(k)}$, and submultiplicativity gives the corresponding exponential suppression of the moment gap under repeated quenches.

\paragraph{Unitary invariance of the spectral norm}
\label{app:unitary_norm}

In deriving Eq.~\eqref{eq:spectral_gap_proj} in the main text, we use the fact that left multiplication by $\calU_0^{k,k}$ preserves the spectral norm. We state this property explicitly below.

\begin{theorem}
\label{thm:unitary_norm}
For any $U_0\in U(d)$ and $A\in\linOpCdk{2}$,
\begin{equation}
    \|\calU_0^{k,k}A\|_\infty
    =
    \|A\|_\infty .
\end{equation}
\end{theorem}

\begin{proof}
Since $\calU_0^{k,k}$ is unitary,
\begin{align}
    \|\calU_0^{k,k}A\|_\infty^2
    &=
    \left\|
    A^\dagger
    \calU_0^{\dagger k,k}
    \calU_0^{k,k}
    A
    \right\|_\infty
    =
    \|A^\dagger A\|_\infty
    =
    \|A\|_\infty^2.
\end{align}
Taking the positive square root gives
$\|\calU_0^{k,k}A\|_\infty=\|A\|_\infty$.
\end{proof}

Finally, we comment on the perturbative treatment used in the weak-disorder regime. Reorganizing the expansion through a Magnus expansion does not improve the short-time accuracy of the second-order kernel $\Omega^{(k)}$. In particular, the resummed
expression $\left\|
    \Mk{\mu_H,\perp}
    e^{-\sigma^2\Omega^{(k)}}
    \Mk{\mu_H,\perp}
    \right\|_\infty$
does not provide a tighter short-time bound than
Eq.~\eqref{eq:gap_smallsigma}. We therefore use the direct second-order perturbative expansion in the main text.

	\section{Energy level statistics of the disordered XXZ chain}\label{app:level_stats}
\begin{figure*}[ht]
		\centering
		\includegraphics[width=0.75\textwidth]{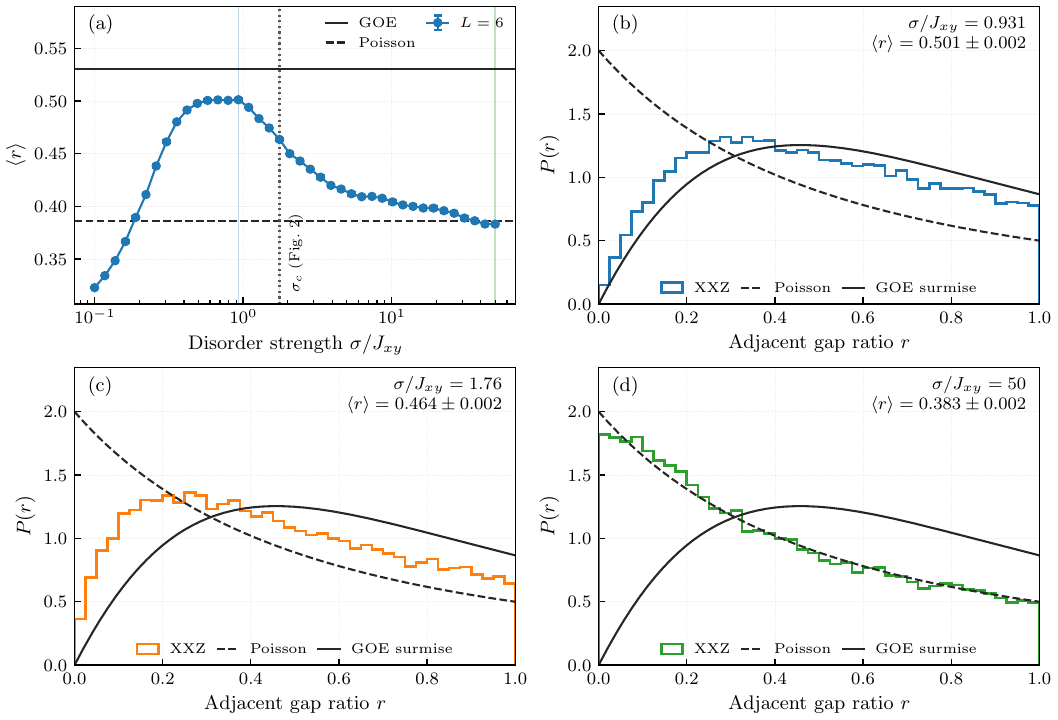}
		\caption{Adjacent-gap statistics of the open XXZ chain with $L=6$, $J_{xy}=1$, $J_z=0.6$, and $S^z=0$, using the Pauli convention of Eq.~\eqref{eq:HXXZ}. Statistics are evaluated over the central half of the spectrum for $5000$ disorder realizations at each disorder strength. (a) Mean adjacent-gap ratio $\langle r\rangle$ with one-standard-error bars. Horizontal lines indicate the Gaussian orthogonal ensemble (GOE, solid) and Poisson (dashed) reference values. The dotted vertical line marks the moment-gap optimum $\sigma_c\simeq 1.7607$. (b)--(d) Distributions $P(r)$ at the sampled maximum of $\langle r\rangle$, $\sigma/J_{xy}\simeq 0.9308$, at $\sigma_c$, and at $\sigma/J_{xy}=\textcolor{red}{50}$, respectively. Solid and dashed curves show the GOE surmise and Poisson distribution. The level-statistics maximum and the moment-gap minimum occur at distinct disorder strengths on this finite chain.}
		\label{fig:level_statis_pr}
	\end{figure*}

    In this section, we compute adjacent-gap statistics for the same XXZ setting and disorder grid as Fig.~\ref{fig:gap_vs_sigma}: $L=6$, $J_{xy}=1$, $J_z=0.6$, Gaussian longitudinal fields, open boundaries, and $S^z=0$. We use the Pauli convention of Eq.~\eqref{eq:HXXZ}, as for all numerical figures.

    From the central half of each spectrum we form
    \begin{equation}
        r_n=\frac{\min(\delta_n,\delta_{n+1})}{\max(\delta_n,\delta_{n+1})},
        \qquad \delta_n=E_{n+1}-E_n.
    \end{equation}
    We average over $5000$ independent realizations at each of $40$ disorder strengths, $0.1\leq\sigma/J_{xy}\leq 50$, with standard errors computed from the per-realization means. This diagnostic requires no spectral unfolding. The reference means are $\langle r\rangle_{\mathrm P}=2\ln2-1\simeq0.3863$ for Poisson statistics and $\langle r\rangle_{\mathrm{GOE}}\simeq0.5307$ for large GOE matrices, and the distribution curves use the GOE surmise~\cite{PhysRevLett.110.084101}.

    Fig.~\ref{fig:level_statis_pr} shows a broad maximum near $0.49\lesssim\sigma/J_{xy}\lesssim 0.93$, with a sampled peak $\langle r\rangle=0.5012\pm0.0015$ at $\sigma/J_{xy}\simeq 0.9308$, followed by a decrease toward Poisson statistics at strong disorder~\cite{Kudo2004XXZStatistics,MBL1}. At the moment-gap minimum $\sigma_c\simeq 1.7607$ from Fig.~\ref{fig:gap_vs_sigma}, $\langle r\rangle=0.4636\pm0.0016$. Thus, the two diagnostics have distinct extrema on this finite chain, and energy-level statistics alone do not establish a unitary $k$-design.

\end{document}